\documentclass[preprint,pre,floats,aps,amsmath,amssymb]{revtex4}

\usepackage{graphicx}
\usepackage{bm}

\usepackage{mathpazo}  
\usepackage{setspace}
\usepackage{multirow}
\usepackage{booktabs}
\usepackage{rotating} 

\begin{document}

\title{A revisit of analytical solutions for the tidal circulation in an idealized, frictional, elongated, rotating, and enclosed basin}
\author{Sang In Won$^{1,*}$, Jinseong Kwon$^{2,*}$, and Sung Yong Kim$^{1,\dag}$}
\affiliation{\footnotesize{$^{1}$Environmental Fluid Mechanics Laboratory, Department of Mechanical Engineering, Korea Advanced Institute of Science and Technology, 291 Daehak-ro, Yuseong-gu, Daejeon 34141, Republic of Korea\\ $^{2}$K-Water Research Institute, Water Energy and Infrastructure Research Center, 125, Yuseong-daero 1689 beon-gil, Yuseong-gu, Daejeon 34045, Republic of Korea\\$^{*}$Both authors contributed equally to this work\\$^{\dag}$Corresponding author: syongkim@kaist.ac.kr}}

\date{\today}

\begin{abstract}
We revisited the analytical solutions for the tidal circulation in an idealized, frictional, elongated, rotating, and enclosed basin. The parameters in the tidal model with a single vertical layer and two vertical layers, including the ratio of basin width to
length and the relative position of the upper
layer, are corrected and modified from \cite{winant.cd.07.1,winant.cd.10.1}.
\end{abstract}

\maketitle
\section{Introduction}

Periodic tidal elevations generate significant tidal currents over the continental shelves and in channels. As one of the classic research topics in coastal oceanography and engineering, the circulation in an enclosed basin has drawn attention with respect to geophysical fluids and their complexity, including circulations associated with density gradients between fresh water and sea water and relevant mixing along with tides [e.g., \cite{taylor.gi.22.1,chao.sy.88.1,odonnell.j.90.1,winant.cd.07.1,mccabe.rm.09.1,winant.cd.10.1}].

The analytical solutions for the tidal circulation in an idealized, frictional, elongated, rotating, and enclosed basin, which has constant width and length and varying depth, have been investigated with the conservation of mass and momentum for the primary tidal constituents [e.g., \cite{taylor.gi.22.1,winant.cd.07.1,winant.cd.10.1}]. A single-layer model and a two-layer model with linearized momentum equations simulate the barotropic and baroclinic tidal currents, respectively.  Although these solutions are derived from an idealized model, they are instructive tools to understand the circulation and relevant dynamics.

In this paper, we summarize the tidal circulation in an enclosed basin using idealized single-layer and two-layer tidal models along with minor corrections and modifications from \cite{winant.cd.07.1,winant.cd.10.1}.

%
%


\section{An idealized model and analytical solutions}

\subsection{Model configuration}

The circulation in a frictional, elongated, rotating, and enclosed basin is revisited with an idealized two-layer model forced by oscillating sea surface heights (e.g., tides) with minor corrections and modifications from \cite{winant.cd.07.1,winant.cd.10.1}. The basin has dimensions of length $L^{*}$, width $2B^{*}$, and maximum depth $H^{*}$ ($H^{*} \ll L^{*}, 2B^{*}$).  The dimensional linearized momentum equations in the channel ($x$) and lateral ($y$) directions of the basin under the pressure gradients and stress divergence are given by,

\begin{align}
     \frac{\partial u^{*}}{\partial t^{*}} - f^{*}v^{*} = -g^{*}\frac{\partial \eta^{*}}{\partial x^{*}} + \kappa^{*}\frac{\partial^{2} u^{*}}{\partial z^{*{^2}}}, \label{eqn:xdir}\\
     \frac{\partial v^{*}}{\partial t^{*}} + f^{*}u^{*} = -g^{*}\frac{\partial \eta^{*}}{\partial y^{*}} + \kappa^{*}\frac{\partial^{2} v^{*}}{\partial z^{*{^2}}}\label{eqn:ydir},
\end{align}
where $g^{*}$ and $\kappa^{*}$ denote the gravitational acceleration and dimensional vertical eddy diffusivity, respectively. Note that dimensional and non-dimensional variables are denoted with and without asterisks, respectively. The momentum equations (Eqs. \ref{eqn:xdir} and \ref{eqn:ydir}) are non-dimensionalized with

\begin{align}
t&=\sigma^* t^*,\\
x&=\frac{x^*}{L^*},\\
y&=\frac{y^*}{B^*},\\
z,h&=\frac{z^*,h^*}{H^*},\\
\sigma&=\frac{\sigma^* L^*}{ \sqrt{g^* H^*}},\\
 f&=\frac{f^* L^*}{ \sqrt{g^* H^*}},\\
     \bar{y} & = \frac{B^{*}}{L^{*}},\\
    \delta
    & =\sqrt{\frac{2\kappa^{*}}{\sigma^{*}H^{*2}}},
\end{align}
where $\eta^{*}$, $\xi^{*}$, $\sigma^{*}$, $f$, $\bar{y}$, and $\delta$ denote the surface elevation and interface fluctuation at the entrance of the basin ($x = 0$), tidal frequency, non-dimensionalized Coriolis frequency scaled by a ratio of the basin length to the Rossby deformation radius, aspect ratio of the basin, and non-dimensionalized vertical eddy viscosity, respectively.

The non-dimensional baroclinic frequency ($\hat{\sigma}$) is defined with the non-dimensionalized density anomaly ($\rho$) and upper layer thickness  ($\hbar$);

\begin{align}
\hat{\sigma}=\frac{\sigma}{ \sqrt{\rho \hbar (1-\hbar)}},
\end{align}
where

\begin{align}
    \rho &=2 \frac{\rho^*_2-\rho^*_1}{\rho^*_2+\rho^*_1},\\
    \hbar &=\frac{\hbar^{*}}{H^*}, \label{eqn:defhbar}
\end{align}
and the subscripts 1 and 2 indicate the variable at the upper and lower layers, respectively ($\hbar^{*}$ denotes the dimensional upper layer thickness).

Additionally, non-dimensionalized dynamic variables for the velocity components ($u$, $v$, and $w$), surface elevation ($\eta$), and interface fluctuation ($\xi$) are given by,

\begin{align}
u_{1,2} &=\frac{u^*_{1,2}}{\epsilon \sigma^* L^*},\\ v_{1,2} &=\frac{v^*_{1,2}}{\epsilon \sigma^* B^*},\\ w_{1,2} &=\frac{w^*_{1,2}}{\epsilon \sigma^* H^*},\\
\eta,\xi & =\frac{\eta^*,\xi^*}{\epsilon \sigma^2 H^*},
\end{align}
where $\epsilon= {C^{*}}/{H^*}$ denotes the ratio of the tidal amplitude at the basin entrance $(C^{*})$ to the maximum depth. Then, the sea surface elevation is represented with
\begin{align}
     \eta = C^{*}N(y)\mathrm{Re}\left(e^{-i\sigma^{*}t^{*}}\right)
\end{align}


\subsection{Momentum equations }


The non-dimensionalized momentum equations in a two-layer model are formulated as,


\begin{align}
	\left(u_{1}\right)_{ t}-\frac{f}{\sigma}\bar{y} v_1 &=- \eta_{x}+\frac{\delta^2}{2}\left(u_{1}\right)_{zz},\label{eqn:m1}\\
	\left(v_{1}\right)_{t}+\frac{f}{\bar{y} \sigma}u_{1} &=-\frac{1}{\bar{y}^2} \eta_{y}+\frac{\delta^2}{2}\left( v_{1}\right)_{zz},\label{eqn:m2}\\
	\left(u_{2}\right)_{t}-\frac{f}{\sigma}\bar{y} v_2 &=-\left( \eta_{x}+\rho \xi_{x}\right)+\frac{\delta^2}{2}\left(u_{2}\right)_{zz},\label{eqn:m3}\\
	\left(v_{2}\right)_{t}+\frac{f}{\bar{y} \sigma}u_2&=-\frac{1}{\bar{y}^2}\left(\eta_{y} +\rho \xi_{y}\right)+\frac{\delta^2}{2}\left(v_{2}\right)_{zz},\label{eqn:m4}
\end{align}
where the subscripts $t$, $x$, $y$, and $z$ denote a partial derivative with respect to the given variable (see Tables \ref{tbl:comppara1} and \ref{tbl:comppara2} for more details of corrected and modified equations).


%
%


To solve the momentum equations (Eqs. \ref{eqn:m1} to \ref{eqn:m4}) in the complex number domain, the complex amplitudes ($U$, $V$, $N$, and $I$) of the dynamic variables ($u$, $v$, $\eta$, and $\xi$) are introduced as,

\begin{align}
	u_{1,2}&=\mathrm{Re}\left(U_{1,2}e^{-it}\right),\label{eqn:complexconvert3}\\
	v_{1,2}&=\mathrm{Re}\left(V_{1,2}e^{-it}\right),\label{eqn:complexconvert4}\\
	\eta &=\mathrm{Re}\left(Ne^{-it}\right), \label{eqn:complexconvert1}\\
	 \xi &=\mathrm{Re}\left(Ie^{-it}\right).\label{eqn:complexconvert2}
\end{align}
Then, the momentum equations at each layer are simplified as,

\begin{align}
	\left(U_{1}\right)_{zz}+\frac{2i}{\delta^2}U_1+\frac{2f \bar{y}}{\sigma \delta^2}V_1&=\frac{2}{\delta^2}N_x,\\
	\left(V_{1}\right)_{zz}+\frac{2i}{\delta^2}V_1-\frac{2f}{\sigma \delta^2 \bar{y}}U_1&=\frac{2}{\delta^2 \bar{y}^2}N_y,\\
	\left(U_{2}\right)_{zz}+\frac{2i}{\delta^2}U_2+\frac{2f \bar{y}}{\sigma \delta^2}V_2&=\frac{2}{\delta^2}\left(N_x+\rho I_x\right),\\
	\left(V_{2}\right)_{zz}+\frac{2i}{\delta^2}V_2-\frac{2f}{\sigma \delta^2 \bar{y}}U_2&=\frac{2}{\delta^2 \bar{y}^2}\left(N_y+\rho I_y\right).
\end{align}

\subsection{Analytical solutions}

The analytical solutions of the current components ($U$ and $V$) are

\begin{align}
	U&=iq^N N_x - \frac{1}{\bar{y}} r^N N_y + i\rho q^I I_x- \frac{\rho}{\bar{y}} r^I I_y, \label{eq:localvel}\\
	V&=\frac{i}{\bar{y}^2} q^N N_y + \frac{1}{\bar{y}} r^N N_x + \frac{i\rho}{\bar{y}^2} q^I I_y+ \frac{\rho}{\bar{y}} r^I I_x,
\end{align}
where $q$ and $r$ denote friction and rotation, respectively. Two parameters ($c_{+}$ and $c_{-}$) are introduced to simplify the friction and rotation terms at the surface and interface [e.g., \cite{winant.cd.10.1}],

\begin{align}
q^N(z)&=-\frac{\sigma^2}{\sigma^2 - f^2}+\frac{\sigma}{2(\sigma+f)}\frac{\cos{c_{+}z}}{\cos{c_{+}h}}+\frac{\sigma}{2(\sigma-f)}\frac{\cos{c_-z}}{\cos{c_{-}h}},\\
	r^N(z)&=-\frac{\sigma f}{\sigma^2-f^2}-\frac{\sigma}{2(\sigma+f)}\frac{\cos{c_{+}z}}{\cos{c_{+}h}}+\frac{\sigma}{2(\sigma-f)}\frac{\cos{c_-z}}{\cos{c_-h}},\\
	q^I(z)&=\begin{cases}
	            \frac{\sigma \left[1-\cos{c_{+}(h-\hbar)}\right]\cos{c_{+}z}}{2(\sigma+f)\cos{c_{+}h}} + \frac{\sigma \left[1-\cos{c_-(h-\hbar)}\right]\cos{c_-z}}{2(\sigma-f)\cos{c_-h}}, \quad -\hbar \leq z < 0\\
            -\frac{\sigma^2}{\sigma^2 - f^2}+\frac{\sigma\left[\cos{c_{+}z}-\sin{c_{+}\hbar}\sin{c_{+}(z+h)}\right]}{2(\sigma+f)\cos{c_{+}h}}+\frac{\sigma\left[\cos{c_-z}-\sin{c_-\hbar}\sin{c_-(z+h)}\right]}{2(\sigma-f)\cos{c_{+}h}}, \quad - h  \leq z < -\hbar
	            \end{cases}\\
	r^I(z)&=\begin{cases}
	            -\frac{\sigma \left[1-\cos{c_{+}(h-\hbar)}\right]\cos{c_{+}z}}{2(\sigma+f)\cos{c_{+}h}} + \frac{\sigma \left[1-\cos{c_-(h-\hbar)}\right]\cos{c_-z}}{2(\sigma-f)\cos{c_-h}}, \quad -\hbar \leq z < 0\\
	            -\frac{\sigma f}{\sigma^2-f^2}-\frac{\sigma\left[\cos{c_{+}z}-\sin{c_{+}\hbar}\sin{c_{+}(z+h)}\right]}{2(\sigma+f)\cos{c_{+}h}}+\frac{\sigma \left[\cos{c_-z}-\sin{c_-\hbar}\sin{c_-(z+h)}\right]}{2(\sigma-f)\cos{c_{+}h}}, \quad - h  \leq z < -\hbar
	            \end{cases}
\end{align}
where $c_{+}$ and $c_{-}$ denote

\begin{align}
	c_{+}& =\frac{1+i}{\delta}\sqrt{\frac{\sigma+f}{\sigma}},\\
	 c_{-}& =\frac{1+i}{\delta}\sqrt{\frac{\sigma-f}{\sigma}},
\end{align}
respectively.

The vertically integrated transports at each layer are given by,

\begin{align}
	\lfloor U_1\rfloor&=i Q^N_1 N_x - \frac{1}{\bar{y}} R^N_1 N_y + i \rho Q^I_1 I_x - \frac{\rho}{\bar{y}} R^I_1 I_y,\\
	\lfloor V_1\rfloor&= \frac{i}{\bar{y}^2} Q^N_1 N_y + \frac{1}{\bar{y}} R^N_1 N_x + \frac{i\rho}{\bar{y}^2} Q^I_1 I_y + \frac{\rho}{\bar{y}} R^I_1 I_x,\\
	\lfloor U_2\rfloor&=i Q^N_2 N_x - \frac{1}{\bar{y}} R^N_2 N_y + i \rho Q^I_2 I_x - \frac{\rho}{\bar{y}} R^I_2 I_y,\\
	\lfloor V_2\rfloor&= \frac{i}{\bar{y}^2} Q^N_2 N_y + \frac{1}{\bar{y}} R^N_2 N_x + \frac{i\rho }{\bar{y}^2} Q^I_2 I_y + \frac{\rho}{\bar{y}} R^I_2 I_x,
\end{align}
where $\lfloor\cdot\rfloor$ indicates the vertical integration of the given variable, i.e.,

\begin{align}
	\lfloor U_1\rfloor, \lfloor V_1 \rfloor,  \lfloor u_1 \rfloor, \lfloor v_1 \rfloor, Q_{1}, R_{1} &=\int_{-\hbar} ^{0} U, V, u, v, q, r \: \mathrm{d}z, \label{eq:tr1}\quad \\ \lfloor U_2\rfloor, \lfloor V_2 \rfloor,  \lfloor u_2 \rfloor, \lfloor v_2 \rfloor, Q_{2}, R_{2} & =\int_{-h}^{-\hbar} U, V, u, v, q, r \: \mathrm{d}z,\label{eq:tr2}
\end{align}
and

\begin{align}
	Q^N_1&=-\frac{\sigma^2 h_1}{\sigma^2 - f^2}+\frac{\sigma^{3/2}\delta\sin{c_{+}h_1}}{2(1+i)(\sigma+f)^{3/2}\cos{c_{+}h_{1}}} +\frac{\sigma^{3/2}\delta\sin{c_-h_1}}{2(1+i)(\sigma-f)^{3/2}\cos{c_-h_{1}}},\\
	Q^N_2&=-\frac{\sigma^2 h_2}{\sigma^2 - f^2}+\frac{\sigma^{3/2}\delta\left[\sin{c_{+}(h_1+h_2)}-\sin{c_{+}h_1}\right]}{2(1+i)(\sigma+f)^{3/2}\cos{c_{+}(h_1+h_{2})}}+\frac{\sigma^{3/2}\delta\left[\sin{c_-(h_1+h_2)}-\sin{c_-h_1}\right]}{2(1+i)(\sigma-f)^{3/2}\cos{c_-(h_1+h_{2})}},\\
	R^N_1&=-\frac{\sigma f h_1}{\sigma^2 - f^2}-\frac{\sigma^{3/2}\delta\sin{c_{+}h_1}}{2(1+i)(\sigma+f)^{3/2}\cos{c_{+}(h_1+h_2)}}+\frac{\sigma^{3/2}\delta\sin{c_-h_1}}{2(1+i)(\sigma-f)^{3/2}\cos{c_-(h_1+h_2)}},\\
	R^N_2&=-\frac{\sigma f h_2}{\sigma^2 - f^2}-\frac{\sigma^{3/2}\delta\left[\sin{c_{+}(h_1+h_2)}-\sin{c_{+}h_1}\right]}{2(1+i)(\sigma+f)^{3/2}\cos{c_+(h_1+h_{2})}}+\frac{\sigma^{3/2}\delta\left[\sin{c_-(h_1+h_2)}-\sin{c_-h_1}\right]}{2(1+i)(\sigma-f)^{3/2}\cos{c_-(h_1+h_{2})}},\\
Q^I_1&=\frac{\sigma^{3/2}\delta(1-\cos{c_+h_2})\sin{c_+h_1}}{2(1+i)(\sigma+f)^{3/2}\cos{c_+(h_1+h_{2})}}+\frac{\sigma^{3/2}\delta(1-\cos{c_-h_2})\sin{c_-h_1}}{2(1+i)(\sigma-f)^{3/2}\cos{c_-(h_1+h_{2})}},\\
	Q^I_2&=-\frac{\sigma^2 h_2}{\sigma^2 - f^2}+\frac{\sigma^{3/2}\delta\left[\sin{c_+(h_1+h_2)}-2\sin{c_+h_1}+\sin{c_+h_1}\cos{c_+h_2}\right]}{2(1+i)(\sigma+f)^{3/2}\cos{c_+(h_1+h_{2})}}\nonumber\\
&+\frac{\sigma^{3/2}\delta\left[\sin{c_-(h_1+h_2)}-2\sin{c_-h_1}+\sin{c_-h_1}\cos{c_-h_2}\right]}{2(1+i)(\sigma-f)^{3/2}\cos{c_-(h_1+h_{2})}},\\
R^I_1&=-\frac{\sigma^{3/2}\delta(1-\cos{c_+h_2})\sin{c_+h_1}}{2(1+i)(\sigma+f)^{3/2}\cos{c_+(h_1+h_{2})}}+\frac{\sigma^{3/2}\delta(1-\cos{c_-h_2})\sin{c_-h_1}}{2(1+i)(\sigma-f)^{3/2}\cos{c_-(h_1+h_{2})}},\\
	R^I_2&=-\frac{\sigma f h_2}{\sigma^2 - f^2}-\frac{\sigma^{3/2}\delta\left[\sin{c_+(h_1+h_2)}-2\sin{c_+h_1}+\sin{c_+h_1}\cos{c_+h_2}\right]}{2(1+i)(\sigma+f)^{3/2}\cos{c_+(h_1+h_{2})}}\nonumber\\
&+\frac{\sigma^{3/2}\delta\left[\sin{c_-(h_1+h_2)}-2\sin{c_-h_1}+\sin{c_-h_1}\cos{c_-h_2}\right]}{2(1+i)(\sigma-f)^{3/2}\cos{c_-(h_1+h_{2})}},
\end{align}
 $h_1= \mathrm{min}(h,\hbar)$, and $h_2= \mathrm{max}(h-\hbar,0)$.

\subsection{Derivation of coupled partial differential equations (PDEs)}

The surface elevation and interface fluctuation are estimated from two vertically integrated mass conservation equations in the entire water column and lower layer:

\begin{align}
	\sigma^2\eta_{t}+  \lfloor u_{1} \rfloor_{x} + \lfloor u_{2} \rfloor_{x}+ \lfloor v_{1}\rfloor_{y} +\lfloor v_{2} \rfloor_{y} &=0,\\
	\sigma^2  \xi_{t}+ \lfloor u_{2} \rfloor_{x}+ \lfloor v_{2}\rfloor_{y}&=0.
\end{align}

With the complex amplitudes (Eqs. \ref{eqn:complexconvert1} to \ref{eqn:complexconvert4}), these equations can be rewritten as

\begin{align}
	-i\sigma^2 N + \lfloor U_1\rfloor_{x}+\lfloor U_2\rfloor_{x}  +  \lfloor V_1 \rfloor_{y}+ \lfloor V_2 \rfloor_{y}&=0,\\
	-i\sigma^2 I + \lfloor U_2\rfloor_x + \lfloor V_2 \rfloor_y&=0.
\end{align}

Then, $\lfloor U_{1}\rfloor$, $\lfloor U_{2}\rfloor$, $\lfloor V_{1}\rfloor$, and $\lfloor V_{2}\rfloor$ can be eliminated by introducing expressions (Eqs. \ref{eq:tr1} and \ref{eq:tr2}) to obtain a coupled set of partial differential equations (PDEs) for the surface elevation and interface fluctuation:

\begin{align}
	&(Q^N_0 N_x)_x+\frac{1}{\bar{y}^2} (Q^N_0 N_y)_y -\frac{i}{\bar{y}}[(R^N_0 N_x)_y-(R^N_0 N_y)_x]-\sigma^2 N \nonumber\\ &+ \rho\left[(Q^I_0 I_x)_x+ \frac{1}{\bar{y}^2} (Q^I_0 I_y)_y -\frac{i}{\bar{y}}(R^I_0 I_x)_y+\frac{i}{\bar{y}}(R^I_0 I_y)_x\right] =0,\label{eqn:pde1}
\end{align}
where the subscript 0 represents a sum of the given variables in the upper and lower layers (e.g., $Q_{0} = Q_{1} + Q_{2}$). The equation in the lower layer is given by,

\begin{align}
	&(Q^N_2 N_x)_x+\frac{1}{\bar{y}^2} (Q^N_2 N_y)_y -\frac{i}{\bar{y}}[(R^N_2 N_x)_y-(R^N_2 N_y)_x]\nonumber\\&+\rho\left[(Q^I_2 I_x)_x+\frac{1}{\bar{y}^2}(Q^I_2 I_y)_y -\frac{i}{\bar{y}}(R^I_2 I_x)_y+\frac{i}{\bar{y}}(R^I_2 I_y)_x\right]-\sigma^2 I=0.\label{eqn:pde2}
\end{align}

\subsection{Boundary conditions}

The coupled PDEs require boundary conditions for the surface elevation and interface fluctuation as the complex amplitudes,

\begin{align}
	N&=N_{0}(y)  \quad \mathrm{at} \quad x=0,\label{eqn:bc1}\\
	 I_{x}&=0 \quad \mathrm{at} \quad x=0,\\
	N_{x}=I_x&=0 \quad \mathrm{at} \quad x=1,\\
	N_{y}&=0 \quad \mathrm{at} \quad y=\pm1,\\
	N_{y}+\rho I_{y}&=0 \quad \mathrm{at} \quad y=\tilde{y}\label{eqn:bc5},
\end{align}
where $N_{0}(y)$ and $\tilde{y}$ indicate the tidal forcing along the entrance of the bay and locations of the lateral boundary of the interface, respectively.



%
%

In addition, the boundary conditions are given at the interface ($z = -\hbar$; Eq. \ref{eqn:defhbar}),

\begin{align}
	U_{1}&=U_{2},\\
	V_{1}& =V_{2}, \\
	\left(U_{1}\right)_{z}&= \left(U_{2}\right)_{z},\\
	\left(V_{1}\right)_{z}&=\left(V_{2}\right)_{z},
\end{align}
and on the bottom ($z = -h$),

\begin{align}
	U_{2}=V_{2}=0.
\end{align}





\subsection{Corrections and modifications of the tidal model}
%

The idealized model and analytical solution of \cite{winant.cd.07.1,winant.cd.10.1} shares the same governing equations, and the surface elevation can be expanded as;
\begin{align}
(Q_{0}^{N}N_{y})_{y}-i\bar{y} [(R_{0}^{N}N_{x})_{y}-(R_{0}^{N}N_{y})_{x}]+\bar{y}^2[(Q_{0}^{N}N_{x})_x-\sigma^2N]=0.
 \label{eq:conti}
\end{align}

The asymptotic expansion of the surface elevation is given by,

\begin{align}
N&=N^{(0)}+\bar{y}N^{(1)}+\bar{y}^{2}N^{(2)}+...\\
 &=N^{(0)}+\bar{y}N^{(1)}+O(\bar{y}^2).
 \label{eq:expansionN}
\end{align}

The analytical solutions are constituted with zeroth-order [$N^{(0)}$] and first-order [$N^{(1)}$] solutions, which correspond to tidal forcing and rotation associated with the Coriolis force, respectively. The zeroth-order and first-order solutions are substituted to the governing equation (Eq. \ref{eqn:pde1}) under a barotropic condition ($\rho=0$) by multiplying all terms by $\bar{y}^2$ to all terms. The governing equations sorted by the order are given by,


\begin{align}
&Q_{0}^{N}\left(N_{y}^{(0)}\right)_{y}+\left(Q_{0}^{N}\right)_{y}N_{y}^{(0)}+\bar{y} Q_{0}^{N}\left(N_{y}^{(1)}\right)_{y}+\bar{y} \left(Q_{0}^{N}\right)_{y}N_{y}^{(1)}\nonumber\\
&-i\bar{y} \left(R_{0}^{N}\right)_{y}N_{x}^{(0)}-i\bar{y} R_{0}^{N}\left(N_{x}^{(0)}\right)_{y}+i\bar{y} \left(R_{0}^{N}\right)_{x}N_{y}^{(0)}+i\bar{y} R_{0}^{N}\left(N_{y}^{(0)}\right)_{x}\nonumber\\
&-i\bar{y}^2 \left(R_{0}^{N}\right)_{y}N_{x}^{(1)}-i\bar{y}^2 R_{0}^{N}\left(N_{x}^{(1)}\right)_{y}+i\bar{y}^2 \left(R_{0}^{N}\right)_{x}N_{y}^{(1)}+i\bar{y}^2 R_{0}^{N}\left(N_{y}^{(1)}\right)_{x}\nonumber \\
&+\bar{y}^2 \left(Q_{0}^{N}\right)_{x}N_{x}^{(0)}+\bar{y}^2 Q_{0}^{N}\left(N_{x}^{(0)}\right)_{x}+\bar{y}^3 \left(Q_{0}^{N}\right)_{x}N_{x}^{(1)}+\bar{y}^3 Q_{0}^{N}\left(N_{x}^{(1)}\right)_{x}\nonumber\\
&+\bar{y}^2\sigma^2N^{(0)}-\bar{y}^3\sigma^2N^{(1)}=0.\label{eq:high}
\end{align}

The first-order equations are given by,
\begin{align}
&iQ_{0}^{N}N_{y}^{(0)}+i\bar{y}Q_{0}^{N}N_{y}^{(1)}+\bar{y}R_{0}^{N}N_{x}^{(0)}+\bar{y}^2R_{0}^{N}N_{x}^{(1)}=0\:\:\mathrm{at}\:\:y = \pm 1.
\end{align}

Other equations are given by,

\begin{align}
\left(i\bar{y}Q_{0}^{N}N_{y}^{(1)}+\bar{y}R_{0}^{N}N_{x}^{(0)}\right)_y&=0, \\
\bar{y}Q_{0}^{N}N_{y}^{(1)}&=i\bar{y}R_{0}^{N}N_{x}^{(0)}\:\: \mathrm{at}\:\:y = \pm1,\\
N^{(1)}&=iN_{x}^{(0)}\int_{0}^{y}\frac{R_{0}^{N}}{Q_{0}^{N}}dy.
\end{align}

\begin{acknowledgments}

Sang In Won and Sung Yong Kim were supported by grants through the Disaster and Safety Management Institute, Ministry of Public Safety and Security (KCG-01-2017-05) and the Basic Science Research Program through the National Research Foundation (NRF), Ministry of Education (NRF-2017R1D1A1B03028285), Republic of Korea. This work is a part of the Master degree thesis of the first author, and the first two authors equally contributed to this work.

\end{acknowledgments}


%


\begin{thebibliography}{6}
\providecommand{\natexlab}[1]{#1}
\providecommand{\url}[1]{\texttt{#1}}
\renewcommand{\UrlFont}{\rmfamily}
\providecommand{\urlprefix}{URL }
\expandafter\ifx\csname urlstyle\endcsname\relax
  \providecommand{\doi}[1]{doi:\discretionary{}{}{}#1}\else
  \providecommand{\doi}{doi:\discretionary{}{}{}\begingroup
  \urlstyle{rm}\Url}\fi
\providecommand{\eprint}[2][]{\url{#2}}

\bibitem[{Chao(1988)}]{chao.sy.88.1}
Chao, S.-Y., 1988: River-forced estuarine plumes. \textit{J. Phys. Oceanogr.},
  \textbf{18}, 72--88, \doi{10.1175/1520-0485(1988)018<0072:RFEP>2.0.CO;2}.

\bibitem[{McCabe et~al.(2009)McCabe, MacCready,, and Hickey}]{mccabe.rm.09.1}
McCabe, R., P.~MacCready, and B.~Hickey, 2009: Ebb-tide dynamics and spreading
  of a large river plume. \textit{J. Phys. Oceanogr.}, \textbf{39}, 2839--2856,
  \doi{10.1175/2009JPO4061.1}.

\bibitem[{{O'D}onnell(1990)}]{odonnell.j.90.1}
{O'D}onnell, J., 1990: The formation and fate of a river plume: {A} numerical
  model. \textit{J. Phys. Oceanogr.}, \textbf{20}, 551--569,
  \doi{10.1175/1520-0485(1990)020<0551:TFAFOA>2.0.CO;2}.

\bibitem[{Taylor(1922)}]{taylor.gi.22.1}
Taylor, G.~I., 1922: {Tidal Oscillations in Gulfs and Rectangular Basins}.
  \textit{Proc. Lond. Math. Soc.}, \textbf{2~(1)}, 148--181,
  \doi{10.1112/plms/s2-20.1.148}.

\bibitem[{Winant(2007)}]{winant.cd.07.1}
Winant, C.~D., 2007: Three-dimensional tidal flow in an elongated, rotating
  basin. \textit{J. Phys. Oceanogr.}, \textbf{37~(9)}, 2345--2362,
  \doi{10.1175/JPO3122.1}.

\bibitem[{Winant(2010)}]{winant.cd.10.1}
Winant, C.~D., 2010: Two-layer tidal circulation in a frictional, rotating
  basin. \textit{J. Phys. Oceanogr.}, \textbf{40~(6)}, 1390--1404,
  \doi{10.1175/2010JPO4342.1}.

\end{thebibliography}

\clearpage\newpage
\begin{sidewaystable*}[h]
\caption{A comparison of the non-dimensionalized parameters, momentum equations, vertically integrated continuity equation (including periodic solutions), periodic solutions, momentum equations for the periodic solutions, and vertically integrated transport (periodic solutions) in \cite{winant.cd.07.1}, \cite{winant.cd.10.1}, and this manuscript.  Note that subscripts 1 and 2 indicate variables at the upper and lower layers, respectively, and subscripts $t$, $x$, $y$, and $z$ denote a partial derivative with respect to the given variable.\label{tbl:comppara1}}
\centering
\scalebox{0.7}{\begin{tabular}{cccc}
\hline
 Terms & \cite{winant.cd.07.1} & \cite{winant.cd.10.1} & Current manuscript  \\\hline
&  $\begin{aligned}
t=\omega^* t^*, x=\frac{x^*}{L^*}, y=\frac{y^*}{B^*}, \end{aligned}$ & $\begin{aligned}
t=\sigma^* t^*, x=\frac{x^*}{L^*}, y=\frac{y^*}{L^*}, \end{aligned}$ & $\begin{aligned}
t=\sigma^* t^*, x=\frac{x^*}{L^*}, y=\frac{y^*}{B^*}, \end{aligned}$\\
\multirow{2}*{Non-dimensional } & $\begin{aligned}
z,h=\frac{z^*,h^*}{H^*}, \kappa=\frac{\omega^* L^*}{ \sqrt{g^* H^*}},  f=\frac{f^*}{ \omega^*}, \end{aligned}$ & $\begin{aligned}
z,h=\frac{z^*,h^*}{H^*}, \sigma=\frac{\sigma^* L^*}{ \sqrt{g^* H^*}},  f=\frac{f^* L^*}{ \sqrt{g^* H^*}}, \end{aligned}$ & $\begin{aligned}
z,h=\frac{z^*,h^*}{H^*}, \sigma=\frac{\sigma^* L^*}{ \sqrt{g^* H^*}},  f=\frac{f^* L^*}{ \sqrt{g^* H^*}}, \end{aligned}$\\
\multirow{2}*{parameters}& $\begin{aligned}
u=\frac{u^*}{\epsilon \omega^* L^*}, v=\frac{v^*}{\epsilon \omega^* B^*}, w=\frac{w^*}{\epsilon \omega^* H^*},\end{aligned}$ & $\begin{aligned}
u_{1,2} =\frac{u^*_{1,2}}{\epsilon \sigma^* L^*}, v_{1,2} =\frac{v^*_{1,2}}{\epsilon \sigma^* L^*}, w_{1,2} =\frac{w^*_{1,2}}{\epsilon \sigma^* H^*}.\end{aligned}$ &
$\begin{aligned}
u_{1,2} =\frac{u^*_{1,2}}{\epsilon \sigma^* L^*}, v_{1,2} =\frac{v^*_{1,2}}{\epsilon \sigma^* B^*}, w_{1,2} =\frac{w^*_{1,2}}{\epsilon \sigma^* H^*}.\end{aligned}$\\
& $\begin{aligned}
\eta  =\frac{\eta^*}{C^*}, \alpha=\frac{B^*}{L^*}.\end{aligned}$ & $\begin{aligned}
\eta,\zeta  =\frac{\eta^*,\zeta^*}{\epsilon \sigma^2 H^*},\bar{y}=\frac{B^*}{L^*}, \epsilon = \frac{C^{*}}{H^{*}}.\end{aligned}$ &
$\begin{aligned}
\eta,\zeta  =\frac{\eta^*,\zeta^*}{\epsilon \sigma^2 H^*},\bar{y}=\frac{B^*}{L^*}, \epsilon = \frac{C^{*}}{H^{*}}.\end{aligned}$\\\hline
\multirow{2}*{Momentum equations}& $\begin{aligned}
	\frac{\partial u}{\partial t}-f\alpha v &=-\frac{1}{\kappa^2}\frac{\partial \eta}{\partial x}+\frac{\delta^2}{2}\frac{\partial^2 u}{\partial z^2},\\
	\frac{\partial v}{\partial t}+\frac{f}{\alpha}u &=-\frac{1}{\alpha^2\kappa^2}\frac{\partial \eta}{\partial y}+\frac{\delta^2}{2}\frac{\partial^2 v}{\partial z^2}.\\
\end{aligned}$ & $\begin{aligned}
	\frac{\partial u_1}{\partial t}-\frac{f}{\sigma} v_1 &=-\frac{\partial \eta}{\partial x}+\frac{\delta^2}{2}\frac{\partial^2 u_1}{\partial z^2},\\
	\frac{\partial v_1}{\partial t}+\frac{f}{\sigma}u_1 &=-\frac{\partial \eta}{\partial y}+\frac{\delta^2}{2}\frac{\partial^2 v_1}{\partial z^2},\\
	\frac{\partial u_2}{\partial t}-\frac{f}{\sigma} v_2 &=-\left(\frac{\partial \eta}{\partial x}+\rho\frac{\partial \zeta}{\partial x}\right)+\frac{\delta^2}{2}\frac{\partial^2 u_2}{\partial z^2},\\
	\frac{\partial v_2}{\partial t}+\frac{f}{\sigma}u_2&=-\left(\frac{\partial \eta}{\partial y}+\rho\frac{\partial \zeta}{\partial y}\right)+\frac{\delta^2}{2}\frac{\partial^2 v_2}{\partial z^2}.
\end{aligned}$  &
$\begin{aligned}
	\frac{\partial u_1}{\partial t}-\frac{f}{\sigma}\bar{y} v_1 &=-\frac{\partial \eta}{\partial x}+\frac{\delta^2}{2}\frac{\partial^2 u_1}{\partial z^2},\\
	\frac{\partial v_1}{\partial t}+\frac{f}{\bar{y} \sigma}u_1 &=-\frac{1}{\bar{y}^2}\frac{\partial \eta}{\partial y}+\frac{\delta^2}{2}\frac{\partial^2 v_1}{\partial z^2},\\
	\frac{\partial u_2}{\partial t}-\frac{f}{\sigma}\bar{y} v_2 &=-\left(\frac{\partial \eta}{\partial x}+\rho\frac{\partial \zeta}{\partial x}\right)+\frac{\delta^2}{2}\frac{\partial^2 u_2}{\partial z^2},\\
	\frac{\partial v_2}{\partial t}+\frac{f}{\bar{y} \sigma}u_2&=-\frac{1}{\bar{y}^2}\left(\frac{\partial \eta}{\partial y}+\rho\frac{\partial \zeta}{\partial y}\right)+\frac{\delta^2}{2}\frac{\partial^2 v_2}{\partial z^2}.
\end{aligned}$\\\hline
\multirow{2}*{Vertically integrated} & \multirow{2}*{$\begin{aligned}
	\frac{\partial \eta}{\partial t}+\frac{\partial \lfloor u \rfloor }{\partial x}+\frac{\partial \lfloor v\rfloor }{\partial y}&=0,\\-iN + \lfloor U\rfloor_{x} +  \lfloor V\rfloor_{y}&=0.\end{aligned}$} &
\multicolumn{2}{c}{$\begin{aligned}
		\sigma^2\frac{\partial \eta}{\partial t}+\frac{\partial \left( \lfloor u_1 \rfloor + \lfloor u_2 \rfloor \right)}{\partial x}+\frac{\partial \left(\lfloor v_1\rfloor +\lfloor v_2 \rfloor \right)}{\partial y}&=0,
\end{aligned}$}\\
{continuity equations} & & \multicolumn{2}{c}{$\begin{aligned}
	\sigma^2\frac{\partial \zeta}{\partial t}+\frac{\partial \lfloor u_2 \rfloor }{\partial x}+\frac{\partial \lfloor v_2\rfloor }{\partial y}&=0.\end{aligned}$}\\\hline
{Periodic solutions} & {$\begin{aligned}
\eta =\mathrm{Re}\left(Ne^{-it}\right),u=\mathrm{Re}\left(Ue^{-it}\right),v=\mathrm{Re}\left(Ve^{-it}\right)\end{aligned}$} & \multicolumn{2}{c}{$\begin{aligned}
\eta =\mathrm{Re}\left(Ne^{-it}\right),\zeta =\mathrm{Re}\left(Ie^{-it}\right), u_{1,2}=\mathrm{Re}\left(U_{1,2}e^{-it}\right),v_{1,2}=\mathrm{Re}\left(V_{1,2}e^{-it}\right)\end{aligned}$} \\\hline
Momentum equations & $\begin{aligned}
	\left(U\right)_{zz}+\frac{2i}{\delta^2}U+\frac{2f \alpha}{\delta^2}V&=\frac{2}{\delta^2 \kappa^2}N_x,\\
	\left(V\right)_{zz}+\frac{2i}{\delta^2}V-\frac{2f}{\alpha \delta^2 }U&=\frac{2}{\delta^2 \alpha^2 \kappa^2}N_y.\end{aligned}$ &
$\begin{aligned}
    \left(U_{1}\right)_{zz}+\frac{2i}{\delta^2}U_1+\frac{2f}{\sigma \delta^2}V_1&=\frac{2}{\delta^2}N_x,\\
	\left(V_{1}\right)_{zz}+\frac{2i}{\delta^2}V_1-\frac{2f}{\sigma \delta^2 }U_1&=\frac{2}{\delta^2 }N_y,\\
	\left(U_{2}\right)_{zz}+\frac{2i}{\delta^2}U_2+\frac{2f }{\sigma \delta^2}V_2&=\frac{2}{\delta^2}\left(N_x+\rho I_x\right),\\
	\left(V_{2}\right)_{zz}+\frac{2i}{\delta^2}V_2-\frac{2f}{\sigma \delta^2 }U_2&=\frac{2}{\delta^2 }\left(N_y+\rho I_y\right).\end{aligned}$ &
$\begin{aligned}
	\left(U_{1}\right)_{zz}+\frac{2i}{\delta^2}U_1+\frac{2f \bar{y}}{\sigma \delta^2}V_1&=\frac{2}{\delta^2}N_x,\\
	\left(V_{1}\right)_{zz}+\frac{2i}{\delta^2}V_1-\frac{2f}{\sigma \delta^2 \bar{y}}U_1&=\frac{2}{\delta^2 \bar{y}^2}N_y,\\
	\left(U_{2}\right)_{zz}+\frac{2i}{\delta^2}U_2+\frac{2f \bar{y}}{\sigma \delta^2}V_2&=\frac{2}{\delta^2}\left(N_x+\rho I_x\right),\\
	\left(V_{2}\right)_{zz}+\frac{2i}{\delta^2}V_2-\frac{2f}{\sigma \delta^2 \bar{y}}U_2&=\frac{2}{\delta^2 \bar{y}^2}\left(N_y+\rho I_y\right).\end{aligned}$\\\hline
\multirow{2}*{Vertically integrated transport} &  $\begin{aligned}
	\lfloor U\rfloor&= \frac{i}{\kappa^2}P_0 N_x - \frac{fQ_0 N_y}{\kappa^2\alpha},\\
	\lfloor V\rfloor&= \frac{i}{\kappa^2\alpha^2} P_0 N_y + \frac{1}{\alpha \kappa^2} fQ_0 N_x. \end{aligned}$ & $\begin{aligned}
	\lfloor U_1\rfloor&=i Q^N_1 N_x - R^N_1 N_y + i \rho Q^I_1 I_x - \rho R^I_1 I_y,\\
	\lfloor V_1\rfloor&= i Q^N_1 N_y +  R^N_1 N_x + i\rho Q^I_1 I_y + \rho R^I_1 I_x,\\
	\lfloor U_2\rfloor&=i Q^N_2 N_x - R^N_2 N_y + i \rho Q^I_2 I_x - \rho R^I_2 I_y,\\
	\lfloor V_2\rfloor&= i Q^N_2 N_y + R^N_2 N_x + i\rho  Q^I_2 I_y + \rho R^I_2 I_x.\end{aligned}$ & $\begin{aligned}
	\lfloor U_1\rfloor&=i Q^N_1 N_x - \frac{1}{\bar{y}} R^N_1 N_y + i \rho Q^I_1 I_x - \frac{\rho}{\bar{y}} R^I_1 I_y,\\
	\lfloor V_1\rfloor&= \frac{i}{\bar{y}^2} Q^N_1 N_y + \frac{1}{\bar{y}} R^N_1 N_x + \frac{i\rho}{\bar{y}^2} Q^I_1 I_y + \frac{\rho}{\bar{y}} R^I_1 I_x,\\
	\lfloor U_2\rfloor&=i Q^N_2 N_x - \frac{1}{\bar{y}} R^N_2 N_y + i \rho Q^I_2 I_x - \frac{\rho}{\bar{y}} R^I_2 I_y,\\
	\lfloor V_2\rfloor&= \frac{i}{\bar{y}^2} Q^N_2 N_y + \frac{1}{\bar{y}} R^N_2 N_x + \frac{i\rho }{\bar{y}^2} Q^I_2 I_y + \frac{\rho}{\bar{y}} R^I_2 I_x.\end{aligned}$\\\hline
\end{tabular}}
\end{sidewaystable*}
\suppressfloats

\clearpage\newpage
\begin{sidewaystable*}[h]
\caption{A comparison of the analytical parameters and vertically integrated continuity equation (periodic solutions) in \cite{winant.cd.07.1}, \cite{winant.cd.10.1}, and this manuscript. Note that subscripts 1 and 2 indicate variables at the upper and lower layers, respectively, and subscripts $t$, $x$, $y$, and $z$ denote a partial derivative with respect to the given variable. \label{tbl:comppara2}}
\centering
\scalebox{0.7}{\begin{tabular}{cccc}
\hline
 Terms & \cite{winant.cd.07.1} & \cite{winant.cd.10.1} & Current manuscript  \\\hline
Analytical parameters &
$\begin{aligned}
	P_0&=\frac{h}{f^2-1}+\frac{\delta \tan{[(1+i)\sqrt{1+f}h/\delta]}}{2(1+i)(1+f)^{3/2}}\\&+\frac{\delta \tan{[(1+i)\sqrt{1-f}h/\delta]}}{2(1+i)(1-f)^{3/2}},\\
    fQ_0&=\frac{fh}{f^2-1}+\frac{\delta \tan{[(1+i)\sqrt{1+f}h/\delta]}}{2(1+i)(1+f)^{3/2}}\\&+\frac{\delta \tan{[(1+i)\sqrt{1-f}h/\delta]}}{2(1+i)(1-f)^{3/2}}.\end{aligned}$ &\multicolumn{2}{c}{$\begin{aligned}
	Q^N_1&=-\frac{\sigma^2 h_1}{\sigma^2 - f^2}+\frac{\sigma^{3/2}\delta\sin{c_+h_1}}{2(1+i)(\sigma+f)^{3/2}\cos{c_+h_{1}}}+\frac{\sigma^{3/2}\delta\sin{c_-h_1}}{2(1+i)(\sigma-f)^{3/2}\cos{c_-h_{1}}},\\
	Q^N_2&=-\frac{\sigma^2 h_2}{\sigma^2 - f^2}+\frac{\sigma^{3/2}\delta\left[\sin{c_+(h_1+h_2)}-\sin{c_+h_1}\right]}{2(1+i)(\sigma+f)^{3/2}\cos{c_+(h_1+h_{2})}}+\frac{\sigma^{3/2}\delta\left[\sin{c_-(h_1+h_2)}-\sin{c_-h_1}\right]}{2(1+i)(\sigma-f)^{3/2}\cos{c_-(h_1+h_{2})}},\\
	R^N_1&=-\frac{\sigma f h_1}{\sigma^2 - f^2}-\frac{\sigma^{3/2}\delta\sin{c_+h_1}}{2(1+i)(\sigma+f)^{3/2}\cos{c_+(h_1+h_2)}}+\frac{\sigma^{3/2}\delta\sin{c_-h_1}}{2(1+i)(\sigma-f)^{3/2}\cos{c_-(h_1+h_2)}},\\
	R^N_2&=-\frac{\sigma f h_2}{\sigma^2 - f^2}-\frac{\sigma^{3/2}\delta\left[\sin{c_+(h_1+h_2)}-\sin{c_+h_1}\right]}{2(1+i)(\sigma+f)^{3/2}\cos{c_+(h_1+h_{2})}}+\frac{\sigma^{3/2}\delta\left[\sin{c_-(h_1+h_2)}-\sin{c_-h_1}\right]}{2(1+i)(\sigma-f)^{3/2}\cos{c_-(h_1+h_{2})}},\\
	Q^I_1&=\frac{\sigma^{3/2}\delta(1-\cos{c_+h_2})\sin{c_+h_1}}{2(1+i)(\sigma+f)^{3/2}\cos{c_+(h_1+h_{2})}}+\frac{\sigma^{3/2}\delta(1-\cos{c_-h_2})\sin{c_-h_1}}{2(1+i)(\sigma-f)^{3/2}\cos{c_-(h_1+h_{2})}},\\
	Q^I_2&=\frac{\sigma^2 h_2}{f^2-\sigma^2}+\frac{\sigma^{3/2}\delta\left[\sin{c_+(h_1+h_2)}-2\sin{c_+h_1}+\sin{c_+h_1}\cos{c_+h_2}\right]}{2(1+i)(\sigma+f)^{3/2}\cos{c_+(h_1+h_{2})}}
\\&+\frac{\sigma^{3/2}\delta\left[\sin{c_-(h_1+h_2)}-2\sin{c_-h_1}+\sin{c_-h_1}\cos{c_-h_2}\right]}{2(1+i)(\sigma-f)^{3/2}\cos{c_-(h_1+h_{2})}},
	\\
R^I_1&=-\frac{\sigma^{3/2}\delta(1-\cos{c_+h_2})\sin{c_+h_1}}{2(1+i)(\sigma+f)^{3/2}\cos{c_+(h_1+h_{2})}}+\frac{\sigma^{3/2}\delta(1-\cos{c_-h_2})\sin{c_-h_1}}{2(1+i)(\sigma-f)^{3/2}\cos{c_-(h_1+h_{2})}},\\
	R^I_2&=-\frac{\sigma f h_2}{\sigma^2-f^2}-\frac{\sigma^{3/2}\delta\left[\sin{c_+(h_1+h_2)}-2\sin{c_+h_1}+\sin{c_+h_1}\cos{c_+h_2}\right]}{2(1+i)(\sigma+f)^{3/2}\cos{c_+(h_1+h_{2})}}\\
	&+\frac{\sigma^{3/2}\delta\left[\sin{c_-(h_1+h_2)}-2\sin{c_-h_1}+\sin{c_-h_1}\cos{c_-h_2}\right]}{2(1+i)(\sigma-f)^{3/2}\cos{c_-(h_1+h_{2})}}, \\ & \mathrm{where}\:\: c_{+} =\frac{1+i}{\delta}\sqrt{\frac{\sigma+f}{\sigma}}\:\: \mathrm{and}\:\:	 c_{-} =\frac{1+i}{\delta}\sqrt{\frac{\sigma-f}{\sigma}}.\\\end{aligned}$}\\\hline
\multirow{3}*{Vertically integrated} & \multirow{4}*{$\begin{aligned}
		&(P_0N_y)_y-i\alpha f[(Q_0 N_x)_y-(Q_0 N_y)_x]\\&+\alpha^2[(P_0N_x)_x-\kappa^2N]=0.\end{aligned}$} & $\begin{aligned}
	(Q^N_0 N_x)_x+ (Q^N_0 N_y)_y -i[(R^N_0 N_x)_y-(R^N_0 N_y)_x]-\sigma^2 N\\+\rho\left[(Q^I_0 I_x)_x+  (Q^I_0 I_y)_y -i(R^I_0 I_x)_y+i(R^I_0 I_y)_x\right]=0,\end{aligned}$ &  $\begin{aligned}
	(Q^N_0 N_x)_x+\frac{1}{\bar{y}^2} (Q^N_0 N_y)_y -\frac{i}{\bar{y}}[(R^N_0 N_x)_y-(R^N_0 N_y)_x]-\sigma^2 N\\+\rho\left[(Q^I_0 I_x)_x+ \frac{1}{\bar{y}^2} (Q^I_0 I_y)_y -\frac{i}{\bar{y}}(R^I_0 I_x)_y+\frac{i}{\bar{y}}(R^I_0 I_y)_x\right]=0,\end{aligned}$\\
\multirow{1}*{continuity equation} & & $\begin{aligned}
		(Q^N_2 N_x)_x+ (Q^N_2 N_y)_y -i[(R^N_2 N_x)_y-(R^N_2 N_y)_x]\\+\rho\left[(Q^I_2 I_x)_x+(Q^I_2 I_y)_y -i(R^I_2 I_x)_y+i(R^I_2 I_y)_x\right]-\sigma^2 I=0.\end{aligned}$ &  $\begin{aligned}
		(Q^N_2 N_x)_x+\frac{1}{\bar{y}^2} (Q^N_2 N_y)_y -\frac{i}{\bar{y}}[(R^N_2 N_x)_y-(R^N_2 N_y)_x]\\+\rho\left[(Q^I_2 I_x)_x+\frac{1}{\bar{y}^2}(Q^I_2 I_y)_y -\frac{i}{\bar{y}}(R^I_2 I_x)_y+\frac{i}{\bar{y}}(R^I_2 I_y)_x\right]-\sigma^2 I=0.\end{aligned}$\\\hline\end{tabular}}
\end{sidewaystable*}
\suppressfloats

\end{document}